\documentclass[twocolumn,prl,showpacs]{revtex4}

\usepackage{graphicx}

\begin{document}

\title{Ratchet-Like Solitonic Transport in Quantum Hall Bilayers}

\author{Ramaz Khomeriki$^1$}
\thanks{khomeriki@hotmail.com}

\author{Lasha Tkeshelashvili$^{2,3}$}
\thanks{lasha@tkm.physik.uni-karlsruhe.de}

\author{Tinatin Buishvili$^1$}

\author{Shota Revishvili$^1$}

\affiliation{$^{1)}$Department of Physics, Tbilisi
State University, Chavchavadze ave.\ 3, Tbilisi 0128, GEORGIA}

\affiliation{$^{2)}$Institut f\"ur Theorie der Kondensierten Materie,
University of Karlsruhe, \\ P.O. Box 6980, 76128 Karlsruhe GERMANY;}

\affiliation{$^{3)}$Institute of Physics, Tamarashvili str.\ 6, Tbilisi 0177, GEORGIA}

\begin{abstract}
\noindent The pseudo-spin model for double layer quantum Hall system with 
total landau level filling factor $\nu=1$ is discussed. Unlike the 
"traditional" one where interlayer voltage enters as static magnetic field 
along pseudo- spin hard axis, in our model we consider applied interlayer 
voltage as a frequency of precessing pseudo-magnetic field lying into the easy 
plane. It is shown that a Landau-Lifshitz equation for the considered pseudo 
magnetic system well describes existing experimental data. Besides that, the 
mentioned model predicts novel directed intra-layer transport phenomenon in 
 the system: unidirectional intra-layer 
energy transport is realized due to interlayer voltage induced motion of 
topological kinks. This effect could be observed experimentally detecting 
counter-propagating intra-layer inhomogeneous charge currents which are 
proportional to the interlayer voltage and total topological charge of the 
pseudo-spin system.

\end{abstract}

\vspace{5mm}


\maketitle

In the recent years there exists a steady interest in quantum Hall bilayers 
with total Landau level filling factor $\nu=1$ due to their anomalous 
transport and tunneling properties \cite{spielman}. Quantum Hall bilayers 
consist of electrons confined in closely separated two dimensional 
semiconductor layers in high magnetic fields. In the 
absence of interlayer voltage,  each layer of the 
system  has a filling factor $\nu_1=\nu_2=1/2$. 
Because of the layers are identical, an electron in 
one layer could be identified with a fake spin up, while 
 in the other layer with a pseudo-spin down. 
Therefore, the system can be 
phenomenologically described via the pseudo-spin formalism \cite{moon}. 
The $z$ component of the  overall pseudo-spin vector specifies the charge 
imbalance between the layers. 
It is clear that the system has a lower energy when the
pseudo-spin points neither up nor down, but rather lies in the plane, 
reflecting the fact that in the ground state  electrons are equally 
distributed between the two layers.
Therefore, within this formalism,  double layer quantum Hall system is 
treated as an easy plane ferromagnet with a  hard axis anisotropy and  
an electron  tunneling between the layers corresponds to the spin flips 
in pseudo-spin language. 

This is a quite satisfactory model for the isolated double layer systems. 
However, 
  problems start to arise as soon as one considers a real experimental 
situation with applied interlayer voltage and induced tunneling current. 
Traditionally, in the phenomenological Hamiltonian, interlayer dc 
voltage is interpreted as a constant magnetic 
field along $z$ axis \cite{moon}. Although this model correctly describes 
the physics at low interlayer voltages \cite{joglekar}, it fails to describe 
experimentally observed current-voltage characteristics \cite{spielman} 
for large interlayer voltages. Besides that, in order to capture the essential  
physics of  the system, one has to introduce effective damping mechanism in 
the formalism. 

Here we use a different interpretation of the  interlayer voltage, 
comprehensive analysis of which can be found in Ref.~\cite{raraki}.   
  Particularly, we treat the interlayer voltage as a circularly 
polarized oscillating magnetic field with the magnitude equal to the tunneling 
amplitude. In addition, it is important that the driving magnetic field 
has the  frequency proportional to the interlayer voltage itself. 
In the absence of damping,  
the traditional model coincides  to our's one 
in the rotating (with the frequency of oscillating field) reference frame. 
Nevertheless, after introducing the damping, the 
 ground state of the system in these two models  differ from each other 
(spins aligned in easy-plane  in our model and tilted from easy-plane 
in the traditional one) and consequently  the equations of motion are no more 
invariant with respect of changing the reference frame. 

 In the equation of motion  we choose Landau-Lifshitz damping term since it 
is most convenient for the systems with easy-plane symmetries. 
The point is that it does not 
lead to the damping to a definite direction in easy-plane. Therefore, 
it is not needed to break the symmetry "by hand" \cite{joglekar}. 
This would not be justified, since the only physically measurable quantity 
is the charge imbalance between the layers, 
i.e. $z$ component of pseudo-spin, while all other components 
of pseudo spin are not physically measurable. As we shall demonstrate below, 
our model well describes experimental observations by Spielman et.al. 
\cite{spielman}. Moreover, this model leads to the prediction of directed 
solitonic transport phenomenon, which can be directly verified on the 
experiment. Note that, this novel effect appears as a natural consequence 
of the identification of interlayer voltage with an ac driving magnetic field. 
Indeed, as it had been suggested recently by Flach and 
co-workers in Ref.~ \cite{flach}, ac driving force with a zero mean value 
may cause a directed  transport in a nonlinear system. 
In addition, it was shown that the necessary condition for this phenomenon 
to appear  is that the system is characterized by nonzero total topological 
charge. As is well established, such topological picture may exist only if the ground state of a system is degenerated, 
what is the case in quantum Hall bilayers according to our model.

Here we present the symmetry analysis \cite{flach}  and  numerical simulations 
  in order to  show that the 
directed inhomogeneous intra-layer current appears due to propagation of 
topological excitations in a quantum Hall bilayer. 
Realistic experimental setup is 
suggested in order to observe this exotic phenomenon.  

{\it The Model:}  The  effective Hamiltonian 
density of our phenomenological model for double layer 
quantum Hall (pseudo) ferromagnet  is given by:
\begin{eqnarray}
{\cal H}=\frac{\rho_E}{2}\left|\frac{\partial m_+}{\partial x}\right|^2
+\beta \bigl(m_z\bigr)^2-\frac{\Delta_{SAS}}{2}\left\{m_+e^{i\omega t}
+\mbox{c.c.}\right\}
\label{1}
\end{eqnarray}
where $\vec m(x,t)$ is a order parameter unit vector; $m_\pm=m_x\pm im_y$ 
($m_z=\nu_1-\nu_2$ describes local electrical charge imbalance between two 
layers); $\rho_E$ is the
in-plane spin stiffness, $\beta$ gives a hard axis anisotropy, $\Delta_{SAS}$ 
is a tunneling amplitude, $\omega=eV/\hbar$ includes interlayer voltage $V$, 
$e$ is an electron charge and "c.c." indicates a complex conjugated term. 
Then Landau-Lifshitz equations of motion,  which  conserve the length of 
a local spin density, can be presented as follows:
\begin{figure}[t]
\begin{center}\leavevmode
\includegraphics[width=0.65\linewidth]{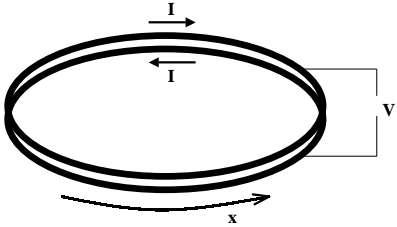}
\vspace{-4.5cm}
\caption{Suggested experimental setup for observation of inhomogeneous 
counter propagating intra-layer charge currents by  
applying an interlayer dc voltage on a quantum Hall bilayer.} \end{center}
\label{ring}
\end{figure}
\begin{equation}
\frac{\partial \vec m}{\partial t}=\left[\vec m\times \vec H\right]-\gamma 
\left[\vec m\times \left[\vec m\times \vec H\right]\right], \label{2}
\end{equation}
where 
\begin{equation}
{\vec H}=-2\left\{\frac{\partial{\cal H}}{\partial{\vec m}}
-\frac{\partial}{\partial x}\left[\frac{\partial{\cal H}}{\partial
\frac{\partial\vec m}{\partial x}}\right]\right\}, \label{3}
\end{equation}
is the effective magnetic field and $\gamma$ is the Landau-Lifshitz 
dimensionless damping coefficient. 

It is easy to see that in the absence of damping our model coincides with the 
traditional one \cite{moon}. Indeed, by redefining the order parameter 
as $m_+\rightarrow m_+\exp(-i\omega t)$, the same equations of 
motion (\ref{2}) could be derived  from the well known 
Hamiltonian density:
\begin{eqnarray}
{\cal H}=\frac{\omega}{2}m_z
+\frac{\rho_E}{2}\left|\frac{\partial m_+}{\partial x}\right|^2
+\beta \bigl(m_z\bigr)^2-\Delta_{SAS}m_x.
\label{4}
\end{eqnarray}
Therefore in case of zero damping the Hamiltonians (\ref{1}) and (\ref{4}) 
describe the same physics in different frames. However, for nonzero damping 
these two approaches are no more equivalent.
\begin{figure}[b]
\begin{center}\leavevmode
\includegraphics[width=0.65\linewidth]{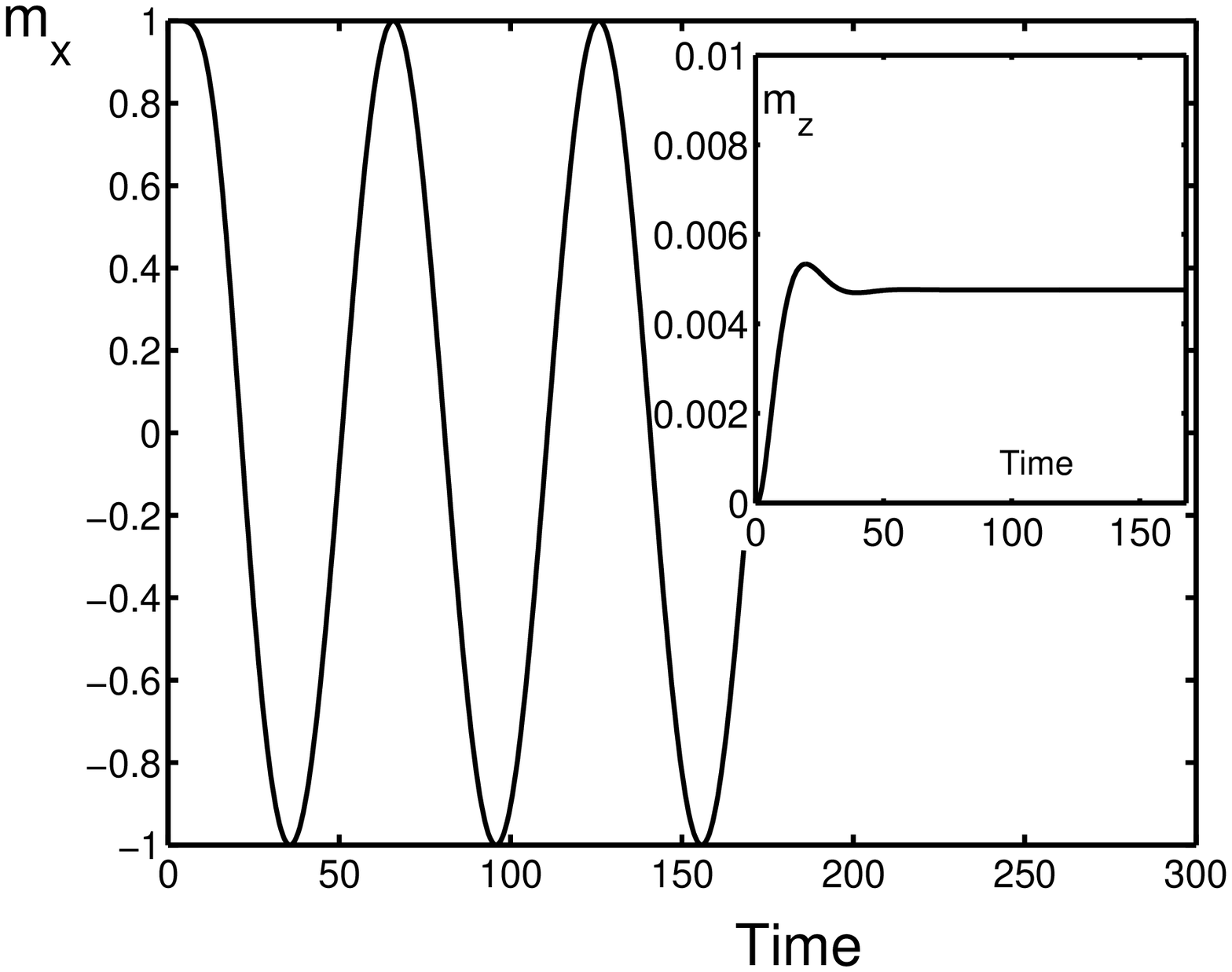}
\includegraphics[width=0.65\linewidth]{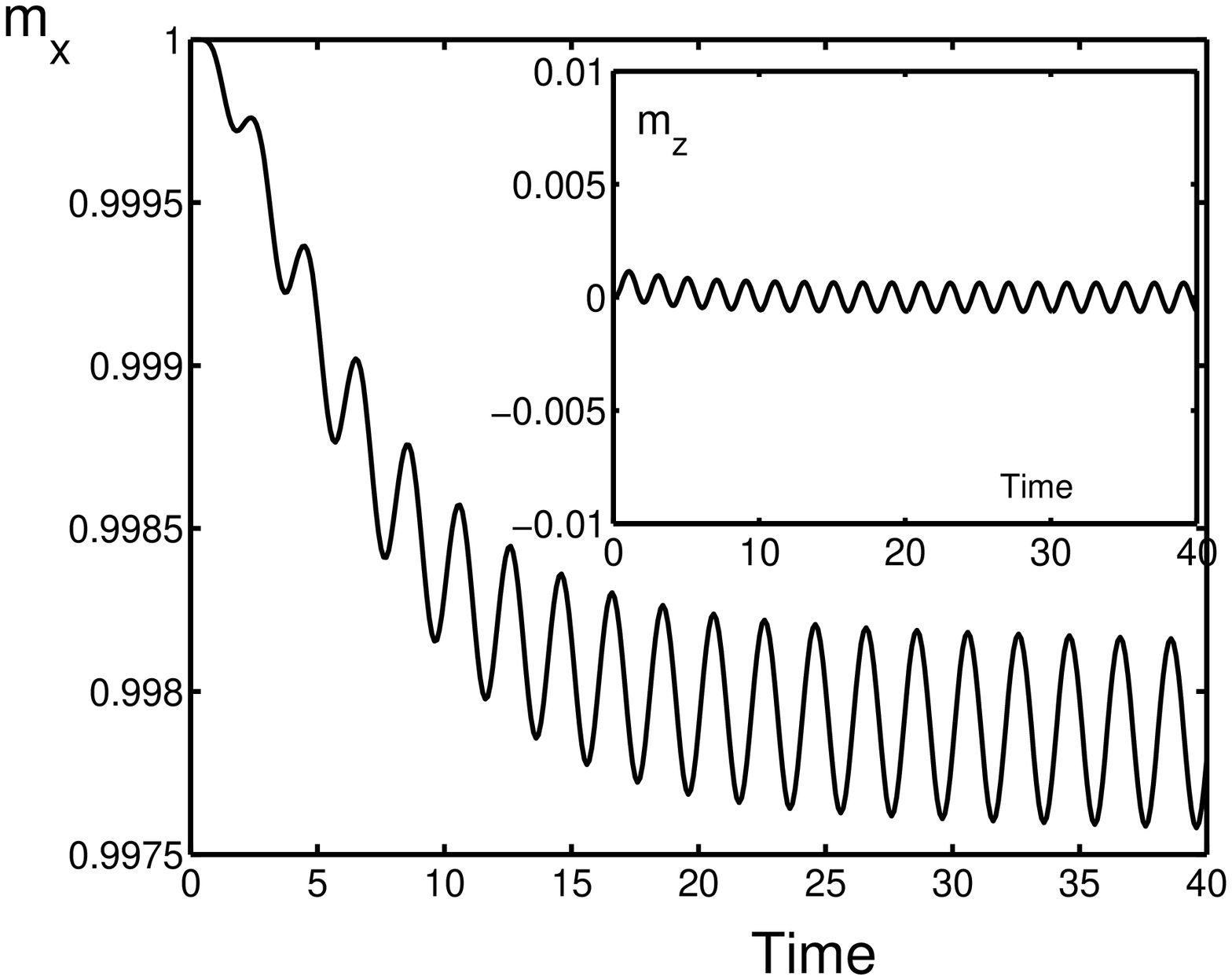}
\caption{Numerical simulation results: pseudo-spin evolution for low 
(upper graph) and large (lower graph) interlayer voltages. Main plots and 
insets represent pseudo-spin $x$ and $z$ components, respectively. Note that 
$m_z$ is small in both low and large voltage limits. In numerical experiments 
it is taken $\beta=10$, $\Delta_{SAS}=0.002$, $\gamma=0.01$, $\omega=0.1$ 
(upper graph) and $\omega=3$ (lower graph).} 
\end{center}
\label{directhall}
\end{figure}

{\it Numerical Experiment and Analytical Solutions:} To specify the problem, 
 we consider 
a quasi-one dimensional quantum Hall double layer system with the 
periodic boundary conditions (quantum Hall "ring" as is shown in Fig. 1). 
By applying the appropriate discretization procedure
$\partial \vec m(x,t)/\partial x \rightarrow \vec m_{i+1}(t)-\vec m_i(t)$ and 
$\partial^2 \vec m(x,t)/\partial x^2 \rightarrow \vec m_{i+1}(t)
+\vec m_{i-1}(t)-2\vec m_i(t)$, the equations of motion  (\ref{2}) 
reduce to a set of $3N$ ordinary differential equations 
($N$ is a number of discretization points). We solve this set of equations 
using MATLAB.  The results are  presented in Fig. 2. 
One can see that the charge imbalance $m_z$ is very small even 
for very large $\omega$-s (comparable with $\beta$).
In the case of small voltages, as is expected, $z$ component of
the order  parameter is slightly tilted from the easy-plane (the 
upper panel in Fig.~2), 
while for the  large voltages, which corresponds to the 
large driving frequencies, $z$ component of  the order
parameter oscillates around the zero value (the lower panel in Fig.~2). 
Such behavior for the large voltages follows from the fact that the driving 
frequency is proportional to the voltage. As is well known \cite{landau}  
a high frequency driving field, even it has a large 
amplitude, can cause only small oscillations in 
a system around the equilibrium point.  
 
 These observations suggest to simplify further the equations of motion. 
Particularly, in the limit $m_z\rightarrow 0$, one can rewrite  
Eq.~(\ref{2})  as follows:
\begin{eqnarray}
4\beta\gamma m_z+\frac{\partial m_z}{\partial t}&=&2\rho_E\frac{\partial^2 
\varphi}{\partial x^2}-2\Delta_{SAS}\sin(\varphi+\omega t) ~~~~~~ \label{5} \\ 
\frac{\partial \varphi}{\partial t}&=&4\beta m_z, \label{6}
\end{eqnarray}
where the phase variable $\varphi$ is defined from the relation $m_+=
\sqrt{1-m_z^2}\exp(i\varphi)$. Reminding that $m_z$ describes a local charge 
imbalance between the layers, Eq.~(\ref{5}) could be interpreted as the charge 
continuity equation with a damping term, where 
\begin{equation}
J_{S}=-2e\rho_E\frac{\partial\varphi}{\partial x}, 
\quad J_{tun}=2e\Delta_{SAS}\sin(\varphi+\omega t) \label{current}
\end{equation}
are intra-layer current in each layer and interlayer tunneling current, 
respectively. Further, substituting (\ref{6}) into the continuity equation 
(\ref{5}), finally we obtain:
\begin{equation}
4\beta\gamma\frac{\partial\varphi}{\partial t}
+\frac{\partial^2\varphi}{\partial t^2}-8\rho_E\beta\frac{\partial^2
\varphi}{\partial x^2}+8\beta\Delta_{SAS}\sin(\varphi+\omega t)=0. \label{7} 
\end{equation}
For the first time, without the gradient term, this equation was suggested 
by Wen and Zee \cite{wen} (see also Ref.~\cite{fogler}).
The heuristical considerations in Ref.~\cite{wen} as well as our numerical 
simulations of  Eq.~(\ref{2}) suggest that, there are completely 
different regimes of the pseudo-spin dynamics for the  
low ($eV/\hbar\ll\sqrt{8\beta\Delta_{SAS}}$) and large 
($eV/\hbar\gg\sqrt{8\beta\Delta_{SAS}}$) voltages. In particular, in the first 
case under the periodic boundary conditions the analytical solution of 
(\ref{7}) should be sought as homogeneous time rotations around $z$ axis 
with the angular frequency $\omega=eV/\hbar$:
\begin{equation}
\varphi=\phi_0-\omega t, \label{8}
\end{equation}
where $\phi_0$ is a constant quantity. Substituting (\ref{8}) into 
the motion equation (\ref{7}) one can define  $\phi_0$ as follows:
\begin{equation}
2\Delta_{SAS}\sin\phi_0=\omega\gamma \label{9}.
\end{equation}
Moreover, from (\ref{current}) the expression for the tunneling current reads:
\begin{equation}
J_{tun}=(e^2/\hbar)V\gamma. \label{10}
\end{equation}

For the case of large voltages the solution is sought as homogeneous 
small time oscillations around a definite direction in the easy plane as:
\begin{equation}
\varphi=A\sin(\phi_0+\omega t), \qquad A\ll 1, \label{11}
\end{equation}
where the constants $A$ and $\phi_0$ must be defined perturbatively by 
substituting (\ref{11}) into the equation of motion (\ref{7}) 
(see Refs.~\cite{fogler,eck}). As a result one obtains:
\begin{equation}
\tan\phi_0=\frac{4\beta\gamma}{\omega}, 
\qquad A=\frac{8\beta\gamma}{\omega\sqrt{\omega^2+16\beta^2\gamma^2}},
 \label{12}
\end{equation}
and the dc component of tunneling current is given by:
\begin{eqnarray}
J_{tun}=2e\Delta_{SAS}\overline{\sin\Bigl[A\sin\bigl(\phi_0+\omega t\bigr)
+\omega t\Bigr]}= \nonumber \\ 
e\Delta_{SAS}A\sin\phi_0
=\frac{e^2}{\hbar}V\frac{32\Delta_{SAS}^2
\beta^2\gamma}{(eV/\hbar)^2\Bigl[(eV/\hbar)^2+16\beta^2\gamma^2\Bigr]}. 
~~~~ \label{13}
\end{eqnarray}
It is easy to see that Eqs.~(\ref{10}) and (\ref{13}) qualitatively well 
describe the experimentally observed tunneling current voltage 
characteristics \cite{spielman}. Indeed, at the low voltages 
the current increases with voltage, while for the large voltages the 
tunneling current decreases as $1/V^3$. 

In the consideration given  above it was assumed that total topological charge of the system is zero.  Consequently we obtain that the  
intra-layer current is zero for any voltages according to the definitions 
(\ref{current}). The situation, however, drastically changes if nonzero topological charge is present in the system.

{\it Directed intra-layer transport:} In order to apply the symmetry analysis 
we should write down the expression for the density of energy flux 
in the system (see e.g. Ref. \cite{flach}):
\begin{equation}
J_E=-\rho_E\frac{\partial\varphi}{\partial x}\frac{\partial\varphi}{\partial t} \label{14}
\end{equation}
and define topological charge of the system as follows: 
\begin{equation}
Q=\Bigl[\varphi(+\infty)-\varphi(-\infty)\Bigr]/2\pi. \label{15}
\end{equation}
In general, the topological charge $Q$ may take any integer value. 
For simplicity in our  numerical experiment we choose an initial 
 topological charge as $Q=1$. That is,  
for the discretized problem of $N$ "spins" we take:
\begin{equation}
m_+^j=exp\Bigl[2i\pi (j-1)/(N+1)\Bigr], \qquad m_z^j=0, \label{16}
\end{equation}
and apply the periodic boundary conditions $\vec m^1=\vec m^{N+1}$ 
to the system. According to the general approach \cite{flach} let us consider 
the following symmetry transformations:
\begin{equation}
x\rightarrow -x, \qquad  \varphi \rightarrow -\varphi. \label{17}
\end{equation}
These symmetry transformations leave the topological charge (\ref{15}) 
invariant but change the sign of the density of energy flux (\ref{14}). 
In this case the symmetry properties of equations of motion becomes crucial. 
 If they are invariant under  the symmetry transformations (\ref{17}) then the 
averaged energy flux in the system is zero, otherwise directed energy flux 
exists even in presence of noise only \cite{flach}. 
The equations of motion (\ref{7}) are not invariant with respect the 
symmetry transform (\ref{17}) and therefore, the directed intra-layer 
energy transport is expected for any applied nonzero interlayer voltage. 
Similarly, for zero voltage Eq.~(\ref{7}) is invariant under the symmetry 
transformation (\ref{17}) and thus averaged energy flux should be zero.

The numerical analysis of both, the initial motion equation (\ref{2}) and its 
reduced version (\ref{7}) completely agrees with the  predictions of the 
symmetry analysis. Thus, in the presence of nonzero total topological charge in the system, directed energy transport should be 
observed. Moreover, the direction of the transport can be changed just by 
inverting the sign of the voltage. 
Besides that, we can further extend the analytical consideration noting that 
the equation of motion (\ref{7}) in the " moving" reference  frame 
$\varphi\rightarrow\varphi-\omega t$ is nothing but dc driven-damped 
sine-Gordon equation (see e.g. Refs.~\cite{malomed,braun}). The role of the 
dc driven force plays the term  $f=4\beta\gamma\omega$. 
As is well known, sine-Gordon equation in the absence of driving force 
supports solitary wave solutions with nonzero topological charge. these 
solutions often are termed as kinks and are given by the following expression: 
\begin{equation}
\varphi=4\arctan\left\{\exp\left[\sqrt{\frac{\Delta_{SAS}}{\rho_E}}(x-x_0)
\right]\right\}. \label{18}
\end{equation}
Applied  driving force results in the motion of that object, velocity of 
which will be proportional to the driving force (applied voltage). Thus, by 
increasing of the voltage, the intra-layer energy transport (and consequently 
the inhomogeneous charge current) will be increased.
\begin{figure}[t]
\begin{center}\leavevmode
\includegraphics[width=0.8\linewidth]{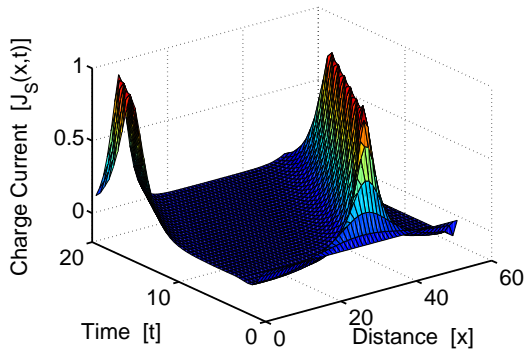}
\caption{Distribution of the  intra-layer local charge current versus time and 
distance in the case of nonzero initial topological charge ($Q=1$) and 
nonzero interlayer voltage. In the numerical experiments the periodic boundary 
conditions are used.} 
\end{center}
\label{dirspinsurf3}
\end{figure}

It should be especially mentioned that if one would  use the  
Hamiltonian (\ref{4}) as a starting point of the analysis, 
one gets the result similar to Eq.~(\ref{7}) but with $\omega=0$. 
Consequently, no directed motion is possible according to the traditional 
model.

{\it Conclusions:}  The driven-damped pseudo-magnetic model is elaborated in 
order to describe the dynamics in quantum Hall bilayers at the total filling 
factor $\nu=1$. The symmetry analysis and numerical simulations  are used 
in order to show that the directed transport exists in  the system 
in the presence of nonzero topological charge. Realistic
experimental setup is suggested for observing  the suggested phenomenon.
Initial nontrivial topological charge [like presented in Exp. (\ref{16})] 
in the system could be realized in the laboratory experiments by application 
of a weak in-plane magnetic field along the double layer "ring" 
($x$ direction in Fig. 1), for which a commensurate pseudo-spin distribution 
appears. Then, switching off the in-plane 
field and applying an interlayer dc voltage it will be possible to observe 
the inhomogeneous counter propagating currents in each layer (see Fig. 3). 
It is obvious that the thermal fluctuations will decrease  the 
topological charge in a  double layer system (like it happens in narrow 
superconducting channels \cite{super}) and as a result the intra-layer 
current eventually should decrease as well.

{\it Acknowledgements:} R. Kh. is indebted to Ramin Abolfath and Kieran Mullen 
for stimulating discussions and is supported by USA Civilian Research and 
Development Foundation award No GP2-2311-TB-02 and NATO reintegration grant 
No PST.EV.979337. L.T.  acknowledges financial support from the Deutsche 
Forschungsgemeinschaft (DFG)  under Bu 1107/2-2 (Emmy-Noether program).

\end{document}